# THE COLOURS OF HII GALAXIES

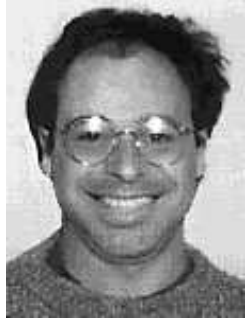

E. TELLES[1,2,3] & R. TERLEVICH[2]
[1] Institute of Astronomy, Madingley Road, Cambridge, CB3 0HA, UK
[2] Royal Greenwich Observatory, Madingley Road, Cambridge, CB3 0EZ, UK

**Abstract**

We present high spatial resolution CCD surface photometry study in the optical V, R and I filters of 15 HII galaxies from the Nordic Optical Telescope and the Jacobus Kapteyn Telescope at Canary Islands. The colours of the starburst continuum and of the underlying galaxy are measured. The distribution of colours of the underlying galaxy in HII galaxies is similar to the colours of other late type low surface brightness galaxies which suggests a close kinship of these with the quiescent phases of HII galaxies. However, comparison with recent evolutionary population synthesis models show that the observational errors and the uncertainties in the models are still too large to put strict constraints on their past star formation history.

## 1 Introduction

The question of whether HII galaxies are primordial galaxies experiencing their very first burst of star formation or if an older stellar population from an earlier event of star formation is present has not, since it was first posed by [8], been answered. Multi-colour surface photometry can provide with the answers. To infer ages and mix of stellar populations of HII galaxies through broad band observations one can compare the observed colours (corrected for the nebular emission line contribution) with those derived from evolutionary synthesis models.

## 2 Comparison of the colours of the underlying galaxies with other dwarf galaxies

Figure 1 shows the result of the comparison of the V-I colours of the *extensions* of HII galaxies (the underlying galaxy) with the total colours of different samples of dwarf galaxies. The underlying galaxy in HII galaxies is best compared with late type low surface brightness (LSB) galaxies (top panels) as opposed to early type dwarfs (e.g. LSB dwarfs ellipticals in Virgo or Fornax). The remarkably blue colours of the underlying galaxy may indicate the lack of an old diffuse red disk such as in high

---

[3]present address: Instituto Astronômico e Geofísico - USP, Caixa Postal 9638, 01065-970 - São Paulo - BRASIL
email: etelles@cosmos.iagusp.usp.br

astro-ph/9511098  21 Nov 1995

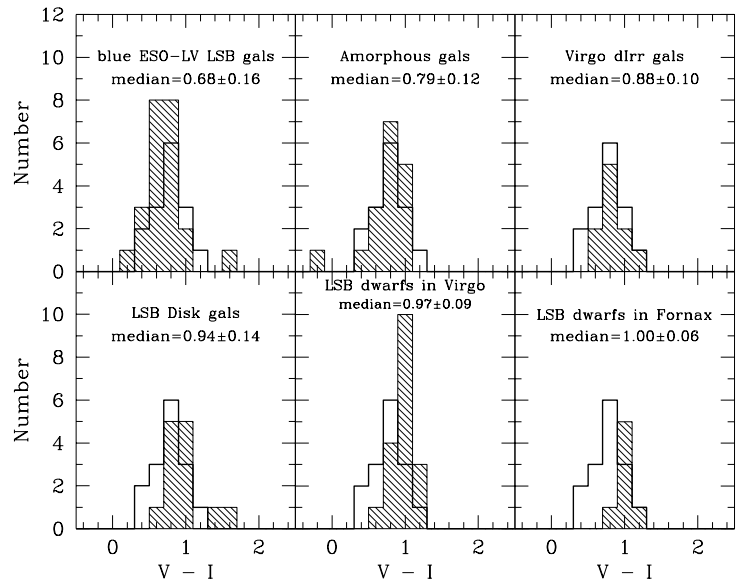

Figure 1: Comparison of the colours of the *extensions* of HII galaxies (solid line histograms) control samples of dwarf galaxies (hatched histograms): Blue ESO-LV LSB galaxies [7]; Amorphous galaxies [4]; Virgo dIrr galaxies [1]; LSB disk galaxies [6]; LSB dwarf ellipticals in Virgo [5]; LSB dwarf ellipticals in Fornax [2]. For HII galaxy these values are V–I = $0.74 \pm 0.14$ and R–I = $0.51 \pm 0.18$.

surface brightness spiral galaxies or extremely low metallicty. Blue LSB galaxies may be quiescent counterparts of HII galaxies in the process of accumulating fuel for the intermittent burst of star formation, until eventual gas depletion. In [9] I also found that "aged" HII galaxies and dwarf ellipticals fall approximately in the same locus in the Luminosity - Surface Brightness diagram and in the size - velocity dispersion diagram. Although the results for the small sample of HII galaxies do not allow us to discuss in detail, the trends are suggestive of the existence the of an evolutionary scenario for dwarf galaxies.

## 3 Comparison of HII galaxy colours with evolutionary population models

### 3.1 The age of the starburst

Figure 2a shows the observed colours of the starburst (filled circles). The colours of the starburst are either redder in V–R or bluer in V–I than the model predictions, even after the emission line subtraction. Also, the colours of the starbursts are still too red as compared to what one expects for a very young single stellar burst ($\approx 10^7$ yrs). Various sources of uncertainties may be the causes of the discrepancies. Some of them are: (i) reddening; (ii) the contribution of the underlying galaxy; (iii) the emission line correction; (iv) uncertainties in the models themselves.

### 3.2 The age of the underlying galaxy

Figure 2b shows the colour-colour diagram for the extensions beyond the ionized region (ext). The position of the galaxy colours in this diagram seems to suggest that the underlying galaxy in HII galaxies have ages ranging from a few $10^9$ to $10^{10}$ years when compared with the SSP models at low metallicity (Z = 1/20 $Z_\odot$). This brief stellar population analysis is consistent with the findings of the comparison with the colours of other dwarf galaxies and seems to rule out the young galaxy hypothesis for these systems.

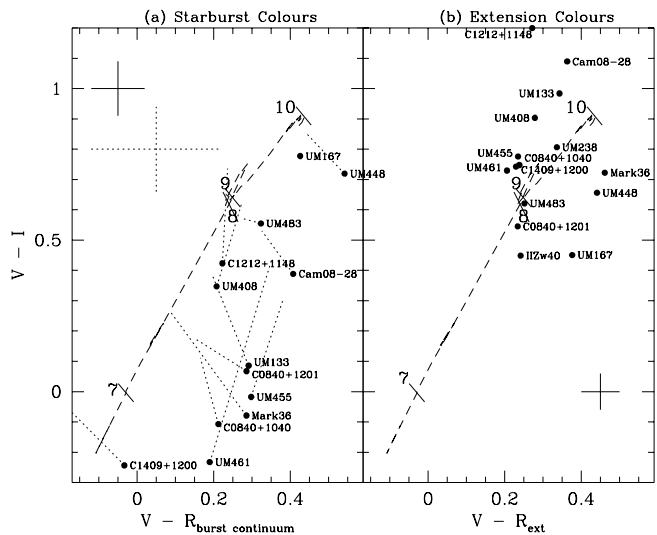

Figure 2: Dereddened colour-colour diagrams. a) Colours of the starburst region alone. The dotted lines drawn from each point illustrate the effect on these colours when the emission line contribution is subtracted. b) Colours of the extensions. The dashed lines are the stellar evolutionary model from [3] for Z=0.001. Ticks are labeled in *log*(age).

## 4 Some of the main conclusions:

(i) In this study, we have confirmed the adequacy of the broad classification scheme, devised in [9]. HII galaxies may be described as two different classes of objects: Type I HII galaxies ($M_V < -18.5$) are more luminous and show disturbed morphology; Type II HII galaxies ($M_V > -18.5$) are compact and regular. They show no signs of being products of interactions or mergers.

(ii) Late type LSB galaxies may be good candidates for being the progenitors of HII galaxies.

(iii) Evolutionary models fail to fit quantitatively the observed colours of the starburst in HII galaxies.

(iv) If the models are right to predict the colours of the intermediate and old stellar population, then the colours of the underlying galaxy in HII galaxies are not compatible with them being truly young galaxies having their very first burst of star formation.

(v) The present analysis does not rule out a possible evolutionary picture for all dwarf galaxies. However, an evolutionary scenario should allow for the different mechanisms which may play a role in triggering the bursting phases in the progenitors, after which these dwarfs could then fade to gas poor systems such as dEs.